# An exact solution to the Dirac equation for a time dependent Hamiltonian in 1-1D space-time.

Dan Solomon
Rauland-Borg Corporation

Email: dan.solomon@rauland.com
May 30, 2009

**Abstract.**

We find an exact solution to the Dirac equation in 1-1 dimension space-time in the presence of a time-dependent potential which consists of a combination of electric, scalar, and pseudoscalar terms.

**1. Introduction.**

In this paper we find an exact solution to the Dirac equation in 1-1 dimensional space-time in the presence of a time-dependent potential which consists of a combination of electric, scalar, and pseudoscalar terms. In the following discussion we set $\hbar = c = 1$ and let $x$ and $t$ stand the space and time dimensions, respectively. In this case the Dirac equation for a single electron in the presence of an external potential $v(x,t)$ is,

$$i\frac{\partial \psi(x,t)}{\partial t} = (H_0 + v(x,t))\psi(x,t) \tag{1.1}$$

where,

$$H_0 = \left(-i\sigma_1 \frac{\partial}{\partial x} + m\sigma_3\right) \tag{1.2}$$

and where $\sigma_j$ ($j=1,2,3$) are the Pauli matrices. In the above $v(x,t)$ must be hermitian. As noted in [1] most general form for $v(x,t)$ is,

$$v(x,t) = V_t(x,t) + \sigma_3 V_s(x,t) + \sigma_1 V_e(x,t) + \sigma_2 V_p(x,t) \tag{1.3}$$

In this expression $V_t$ and $V_e$ stands for the time and space components, respectively, of the 2-vector potential, $V_s$ stands for the scalar term, and $V_p$ is the pseudoscalar term. As discussed in [1], the term $V_e(x,t)$ can always be set equal to zero by a gauge transformation. So in the following discussion we set $V_e(x,t) = 0$. Therefore, using the above, the Dirac equation (1.1) can be written as,

$$i\frac{\partial \psi(x,t)}{\partial t} = (H_0 + V_t(x,t) + \sigma_3 V_s(x,t) + \sigma_2 V_p(x,t))\psi(x,t) \tag{1.4}$$

Some examples of solutions of the Dirac equation in the presence of a static pseudopotential are given in Ref. [1] – [4].

**2. Time dependent potentials.**

Let the potential terms satisfy the following relationships,

$$V_s(x,t) = m(\cos(2g(x,t)) - 1) \tag{2.1}$$

$$V_p(x,t) = -m\sin(2g(x,t)) \tag{2.2}$$

$$V_t(x,t) = \frac{\partial f(x,t)}{\partial t} + \frac{\partial g(x,t)}{\partial x} \tag{2.3}$$

In the above expressions $g(x,t)$ is an arbitrary function and $f(x,t)$ satisfies,

$$\frac{\partial f(x,t)}{\partial x} = -\frac{\partial g(x,t)}{\partial t} \tag{2.4}$$

Given the above relationships it will be shown that the solution to the Dirac equation (1.4) can be written as,

$$\psi(x,t) = e^{-if(x,t)} e^{-i\sigma_1 g(x,t)} \psi_0(x,t) \tag{2.5}$$

where $\psi_0(x,t)$ is the solution to the free field Dirac equation, i.e.,

$$i\frac{\partial \psi_0(x,t)}{\partial t} = H_0 \psi_0(x,t) \tag{2.6}$$

Solutions to this free field equation in 1-1 dimensional space-time are given in [5] (See Eq. 2.4 of [5]).

Define $U = e^{-if(x,t)} e^{-i\sigma_1 g(x,t)}$ so that $\psi(x,t) = U\psi_0(x,t)$. To prove that (2.5) is a solution to (1.4), given (2.1) - (2.4), we will substitute $\psi(x,t) = U\psi_0(x,t)$ into both sides of (1.4) and show that the equality still holds. This yields,

$$i\left(\frac{\partial U}{\partial t}\right)\psi_0 + iU\left(\frac{\partial \psi_0}{\partial t}\right) = \left(H_0 + V_t + \sigma_3 V_s + \sigma_2 V_p\right)U\psi_0 \tag{2.7}$$

Multiply both sides by $U^{-1} = e^{+i\sigma_1 g(x,t)} e^{+if(x,t)}$ to obtain,

$$iU^{-1}\left(\frac{\partial U}{\partial t}\right)\psi_0 + i\left(\frac{\partial \psi_0}{\partial t}\right) = U^{-1}\left(H_0 + V_t + \sigma_3 V_s + \sigma_2 V_p\right)U\psi_0 \tag{2.8}$$

The Pauli matrices satisfy $\{\sigma_i, \sigma_j\} = 2\delta_{ij}$. Using this we obtain the following useful relationships,

$$[U,\sigma_1] = [U^{-1},\sigma_1] = 0; \quad e^{i\sigma_1 g(x,t)}\sigma_3 = \sigma_3 e^{-i\sigma_1 g(x,t)}; \quad e^{i\sigma_1 g(x,t)}\sigma_2 = \sigma_2 e^{-i\sigma_1 g(x,t)} \tag{2.9}$$

From this we obtain,

$$U^{-1}\sigma_3 U = e^{i\sigma_1 g(x,t)} \sigma_3 e^{-i\sigma_1 g(x,t)} = \sigma_3 e^{-i\sigma_1 2g(x,t)} \tag{2.10}$$

and

$$U^{-1}\sigma_2 U = e^{i\sigma_1 g(x,t)} \sigma_2 e^{-i\sigma_1 g(x,t)} = \sigma_2 e^{-i\sigma_1 2g(x,t)} \tag{2.11}$$

To evaluate (2.8) use,

$$U^{-1}H_0(U\psi_0) = \left(-i\sigma_1 U^{-1}\frac{\partial U}{\partial x}\psi_0 - i\sigma_1 \frac{\partial \psi_0}{\partial x} + mU^{-1}\sigma_3 U\psi_0\right) \quad (2.12)$$

Next use,

$$U^{-1}\frac{\partial U}{\partial x} = -i\left(\frac{\partial f(x,t)}{\partial x} + \sigma_1 \frac{\partial g(x,t)}{\partial x}\right) \quad (2.13)$$

along with (2.10) and (2.11) in (2.12) to obtain,

$$U^{-1}H_0(U\psi_0) = -\left(\sigma_1 \frac{\partial f(x,t)}{\partial x} + \frac{\partial g(x,t)}{\partial x}\right)\psi_0 + H_0\psi_0 + m\sigma_3\left(e^{-i\sigma_1 2g} - 1\right)\psi_0 \quad (2.14)$$

We also have that,

$$U^{-1}\left(\frac{\partial U}{\partial t}\right) = -i\left(\frac{\partial f(x,t)}{\partial t} + \sigma_1 \frac{\partial g(x,t)}{\partial t}\right) \quad (2.15)$$

Use this and (2.14) in (2.8) to obtain,

$$\left(\frac{\partial f}{\partial t} + \sigma_1 \frac{\partial g}{\partial t}\right)\psi_0 + i\left(\frac{\partial \psi_0}{\partial t}\right) = -\left(\sigma_1 \frac{\partial f}{\partial x} + \frac{\partial g}{\partial x}\right)\psi_0 + H_0\psi_0 - m\sigma_3\psi_0 + V_t\psi_0 \\ + \left(m\sigma_3 + \sigma_3 V_s + \sigma_2 V_p\right)e^{-i\sigma_1 2g}\psi_0 \quad (2.16)$$

Next, use (2.6) in the above to obtain,

$$\left(\frac{\partial f}{\partial t} + \sigma_1 \frac{\partial g}{\partial t}\right)\psi_0 = -\left(\sigma_1 \frac{\partial f}{\partial x} + \frac{\partial g}{\partial x}\right)\psi_0 - m\sigma_3\psi_0 + V_t\psi_0 + \left(m\sigma_3 + \sigma_3 V_s + \sigma_2 V_p\right)e^{-i\sigma_1 2g}\psi_0$$

$$(2.17)$$

Rearrange terms to obtain,

$$\left(\frac{\partial f}{\partial t} + \sigma_1 \frac{\partial g}{\partial t}\right) + \left(\sigma_1 \frac{\partial f}{\partial x} + \frac{\partial g}{\partial x}\right) + m\sigma_3 - V_t = \left((m + V_s)\sigma_3 + \sigma_2 V_p\right)e^{-i\sigma_1 2g} \quad (2.18)$$

Use (2.3) and (2.4) in the above to obtain,

$$m\sigma_3 e^{i\sigma_1 2g} = (m + V_s)\sigma_3 + \sigma_2 V_p \quad (2.19)$$

Use $\sigma_2 = i\sigma_1\sigma_3$ along with (2.1) and (2.2) in the above,

$$\sigma_3 e^{i\sigma_1 2g} = \sigma_3 \cos(2g) - i\sigma_1\sigma_3 \sin(2g) = \sigma_3\left(\cos(2g) + i\sigma_1 \sin(2g)\right) \quad (2.20)$$

This equality holds because $e^{i\sigma_1 2g} = \cos(2g) + i\sigma_1 \sin(2g)$. Therefore it is proved that (2.5) is a solution of the Dirac equation (1.4) where the potential terms are given by equation (2.1) - (2.4).

Another result that follows is that we can easily obtain exact solutions for the time-independent Hamiltonian which is obtained by making $g$ time independent, i.e., setting $g(x,t) = g(x)$ in Eqs. (2.1) - (2.3) where as shown by (2.4) $f(x,t) = 0$. In this case the solution to the Dirac equation is $\psi(x,t) = e^{-i\sigma_1 g(x)} \psi_0(x,t)$.

**References.**